\documentclass[twocolumn,aps,pra]{revtex4-2} %superscriptaddress,eqsecnum,
\usepackage{amsmath,amsfonts,bm}
\usepackage{graphicx,epstopdf}  %epsfig,
\usepackage{color}
\usepackage{extarrows}
\usepackage{enumitem}
\usepackage{empheq}
\usepackage{tensor}
\newcommand{\ud}{{\mathrm{d}}}
\newcommand{\ue}{{\mathrm{e}}}

\newcommand{\w}{\omega}
\newcommand{\wti}{\widetilde}
\newcommand{\ti}{\tilde}

\newcommand{\B}{\mbox{\tiny B}}

\newcommand{\E}{\mbox{\tiny E}}

\newcommand{\tS}{\mbox{\tiny S}}
\newcommand{\T}{\mbox{\tiny T}}

\newcommand{\dg}{\dagger}
\newcommand{\la}{\langle}
\newcommand{\ra}{\rangle}

\newcommand{\Sec}[1]{Sec.\,\ref{#1}}

\newcommand{\nl}{\nonumber \\}
\newcommand{\be}{\begin{equation}}
\newcommand{\ee}{\end{equation}}
\newcommand{\bsube}{\begin{subequations}}
\newcommand{\esube}{\end{subequations}}
\newcommand{\Eq}[1]{Eq.\,(\ref{#1})}
\newcommand{\Eqs}[1]{Eqs.\,(\ref{#1})}
\newcommand{\Fig}[1]{Fig.\,\ref{#1}}

\newcommand{\RN}[1]{%
  \textup{\uppercase\expandafter{\romannumeral#1}}%
}

\allowdisplaybreaks[1]

\usepackage{hyperref}  
\hypersetup{hidelinks,
	colorlinks=true,
	allcolors=black,
	pdfstartview=Fit,
	breaklinks=true
}

\begin{document}

\title{Quantum Mechanics of Open Systems in Non-Inertial Motion
}
%%%
\author{Zi-Fan Zhu}
% \thanks{Authors of equal contributions}
\author{Yu Su}
\email{suyupilemao@mail.ustc.edu.cn}
% \thanks{Authors of equal contributions}
\author{Yao Wang}
\email{wy2010@ustc.edu.cn}
\author{Rui-Xue Xu}
\email{rxxu@ustc.edu.cn}
\author{YiJing Yan}

\affiliation{Hefei National Research Center for Physical Sciences at the Microscale and Department of Chemical Physics, University of Science and Technology of China, Hefei, Anhui 230026, China}
\affiliation{Hefei National Laboratory, University of Science and Technology of China, Heifei 230088, China}

\date{\today}

\begin{abstract}
The study of quantum mechanics in non-inertial reference frames, particularly in the context of open systems, introduces several intriguing phenomena and challenges.
This paper presents a comprehensive framework for analyzing the quantum mechanics of open systems undergoing non-inertial motion. 
Our methodology leverages the concept of dissipatons, statistical quasi-particles that capture collective dissipative effects from the environment.  We demonstrate that our approach offers a natural understanding of the intricate dynamics among non-inertial effects, decoherence, dissipation, and system-bath entanglement. Specifically, we conduct demonstrations focusing on the Lamb shift phenomenon within a rotating ring cavity. Through theoretical exposition and practical applications, our framework elucidates the profound interplay between open quantum dynamics and non-inertial motion, paving the way for advancements in quantum information processing and sensing technologies.
\end{abstract}
\maketitle

\section{Introduction}

In quantum physics, exploring open quantum systems has become a captivating research frontier \cite{Wei21,Kle09,Bre02,Bre16021002,Dev17015001}.
Open systems, which interact with their surrounding environment, exhibit diverse phenomena spanning quantum optics \cite{Scu97, Lou73, Haa7398, Hak70, Sar74}, nuclear magnetic resonance \cite{Rei82, Sli90, Van051037}, condensed matter \cite{Bor85, Hol59325, Hol59343, Kli97, Ram98}, quark-gluon plasma \cite{Aka15056002, Bla181, Miu20034011, Yao212130010}, nonlinear spectroscopy \cite{Muk95, She84, Muk81509, Yan885160, Yan91179, Che964565, Tan939496, Tan943049}, and chemical and biological physics \cite{Nit06, Lee071462, Eng07782, Dor132746, Cre13253601, Kun22015101}.
The interaction between  open quantum system and its environment often induces decoherence, a process in which the system's quantum coherence diminishes gradually, leading to classical-like behavior.
Decoherence stems from the unavoidable entanglement with the environment, causing a rapid loss of delicate quantum superposition states. 
Consequently, the system is driven towards a mixed state, making it more susceptible to classical statistical treatments.
 In open quantum systems, decoherence plays a vital role in understanding the fundamental limits of quantum technologies and the boundary between classical and quantum behaviors \cite{Max191}.

Non-inertial effects, arising from the acceleration or rotation of the system, introduce novel dynamics and unique features that set them apart from their inertial counterparts \cite{Gho14527}.
Non-inertial effects complicate the dynamics of open quantum systems, including the quantum decoherence. 
 Such effects have garnered significant attention in recent years, primarily due to their potential applications in various fields, including quantum information processing, precision measurements, and quantum metrology \cite{Yan15321,Ahm144996,Li15Arxiv1507_08790,Ary23085011}.

%One prominent example of non-inertial effects in open quantum systems is the Unruh effect \cite{xxx}. This effect, predicted by physicist William G. Unruh in 1976, suggests that an accelerated observer in empty space will perceive a thermal bath of particles, even when the vacuum state is devoid of any particles as observed by an inertial observer. This concept challenges the notion of a unique vacuum state and gives rise to an intriguing connection between quantum field theory and general relativity. The Unruh effect has been successfully observed in several experimental set-ups, demonstrating the profound implications of non-inertial effects on the quantum behavior of open systems \cite{xxx}.

Furthermore, non-inertial effects have also been investigated in the context of quantum entanglement dynamics \cite{Hor09865,Sun1790}.
 Entanglement describes the correlation between two or more distant quantum systems. The acceleration or rotation of an open quantum system can lead to remarkable changes in the entanglement structure between the system and its environment. 
 Understanding these dynamics is crucial in various scenarios such as quantum teleportation, quantum cryptography, and quantum communication protocols \cite{Ren1770,Pir201012,Met14771}.

In this paper, we aim to provide a comprehensive framework for analyzing the quantum mechanics of open systems undergoing non-inertial motion. 
In \Sec{thsec2} and \ref{thsec3}, we will delve into the mathematical formalism necessary to describe the non-inertial effects in open quantum systems, 
followed by a comprehensive framework, leverages the concept of dissipatons, statistical quasi-particles that capture collective dissipative effects from the environment \cite{Yan14054105,Wan22170901}.
% In \Sec{thsec4},
%
 % to elucidate the key ingredients of our formalism, we supply an example of the atom-photon interactions  in a rotating ring cavity, with discussions of the non-inertial effects on the Lamb shift \cite{Ary23085011}.
%
Finally, we provide a concise summary of this paper and deliver a preview of our future research endeavors. 
Throughout this paper, we set $\hbar = c =1$ and $\beta = 1/(k_BT)$, with $k_B$ being the Boltzmann constant and $T$ the temperature. Besides, we denote the space--time coordinate as $x^{\mu} = (t,\bm r) = (t, r^i)$ with $\mu\in\{0,1,2,3\}$ and $i\in\{1,2,3\}$ and the Minkowski metric as $\eta^{\mu\nu} = {\rm diag}\{-1, 1, 1, 1\}$. The Einstein summation convention is also adopted for the space--time indices.

\section{Non-inertial quantum mechanics}\label{thsec2}
\subsection{Non-inertial unitary transformation }
\begin{figure}[h]  
\centering\includegraphics[width=\columnwidth]{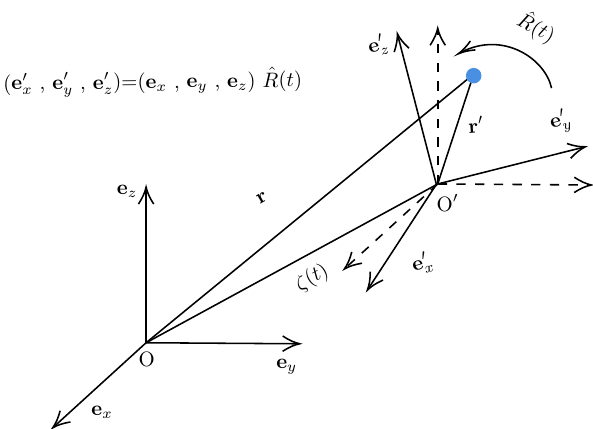} 
\caption{Coordinate transformation of the mixed accelerating and rotational motions.}      
\label{fig1}   
\end{figure}

In this section, we briefly review the non-inertial quantum mechanics \cite{Tak91463}. Let us start with the Hamiltonian,
\begin{align}\label{Hamil}
H(\hat{\bm r},\hat{\bm p})=\frac{\left[\hat{\bm p}-e\bm A(\hat{\bm r},t) \right]^2}{2m}+V(\hat{\bm r},t)+e\phi(\hat{\bm r},t),
\end{align}
where $m$ and $e$ are the mass and charge of the particle, respectively. We use the hat accent $\hat{}$ to distinguish operators and c-numbers for $\bm r$ and $\bm p$. Here, 
$\bm A(\hat{\bm r},t)$ and $\phi(\hat{\bm r},t)$ are the electromagnetic (EM) potentials, while $V(\hat{\bm r},t)$ is the potential energy. All of them possibly contain non-inertial effects as elaborated below. 
To be concrete, we set
\begin{align}\label{potential}
    V(\hat{\bm r},t)=V_0\big(\hat{\bm r}'(t)\big), 
\end{align}
with $\hat{\bm r}'=\mathsf{R}_t^{-1}\left(\hat{\bm r}-{\bm \zeta}_t\right)$, and
 $V_0$ being a specific potential in the body-fixed coordinate system. 
Here, $\mathsf{R}_t \in {\rm SO}(3)$ and $\bm \zeta_t \in \mathbb{R}^3$. By convention, we set the initial conditions: $\mathsf{R}_{0}=\mathsf{I}$, and $\bm\zeta_{0}={ 0}$; See \Fig{fig1} for the illustration.
Special cases include: 
\begin{enumerate}[fullwidth]
    \item Pure acceleration: $\mathsf{R}_t\equiv \mathsf{I}$, $V(\hat{\bm r},t)=V_{0}(\hat{\bm r}-\bm\zeta_t)$. Especially, $\bm\zeta_t=\bm v t$ and $\bm\zeta_t=(1/2)\bm a t^2$ correspond to the boost and the constant acceleration, respectively;
    \item Pure rotation: $ \bm\zeta_t\equiv 0$, $V(\hat{\bm r},t)=V_0(\mathsf{R}_t^{-1}\hat{\bm r})$. Generally, $\mathsf{R}_t$ can be expressed as, 
    \be
\exp_{+}\bigg[\int_{0}^{t}\!\!\ud\tau\ \Omega(\tau){\bf n}(\tau)\cdot{\mathbf J}\bigg].
    \ee
    Here, $\exp_{+}$ denotes the exponential with time-ordering, $\Omega(t)$ is the transient angular velocity, $\hat{\bf n}(t)$ is the unit vector along the rotational direction, and $\mathbf J\equiv(\mathsf J_x, \mathsf J_y, \mathsf J_z)$ with $(\mathsf J_i)_{jk}=-\epsilon_{ijk}$ is the generator of the ${\rm SO}(3)$ group. Especially, $\exp(\Omega t {\mathsf J}_z)$ represents the rotation around the $z$-axis with a constant angular velocity $\Omega$.
\end{enumerate}

Consider the time--dependent unitary transformation
\be
|\psi\rangle\longrightarrow |\psi'\rangle= U(t)|\psi\rangle,
\ee
with 
\begin{align}\label{uni}
    U(t)&=
    \exp_{-}\bigg[i\int_{0}^{t}\!\ud\tau\, \Omega(\tau){\bf n}(\tau)\cdot\hat{\bm L}\bigg]
\exp\bigg(\!-im\!\int_{0}^t\!\ud\tau\,\frac{\dot{\bm{\zeta}}_{\tau}^2}{2}\bigg)
 \nl 
    & \quad
\times 
    \exp\big(-im\dot{\bm{\zeta}}_t\cdot\hat{\bm r}\big)\exp\big(i\bm{\zeta}_t\cdot\hat{\bm p}\big).
\end{align}
Here, $\hat{\bm L}=\hat{\bm r}\times\hat{\bm p}$ is the angular momentum and $\exp_{-}$ denotes the exponential with anti-time-ordering, . The Hamiltonian transforms accordingly as 
\be
H\longrightarrow H'=UHU^{\dagger}+i\dot{U}U^{\dagger}.
\ee
As a result, the new Hamiltonian reads 
\begin{align}\label{Hprime}
H'&=\frac{[\mathsf{R}_t\hat{\bm p}\!-\!e{\bm A}(\hat{\bm x}_t,t)]^2}{2m}
\!+\!V_0(\hat{\bm r})+e[\phi(\hat{\bm x}_t,t)-\!\dot{\bm\zeta}\!\cdot {\bm A}(\hat{\bm x}_t,t)]
\nl
&\quad-\Omega(t){\bf n}(t)\cdot\big[(\mathsf{R}_t\hat{\bm r})\times(\mathsf{R}_t\hat{\bm p})\big]
%
% &\frac{1}{2m}\big[\mathsf{R}_t\hat{\bm p}+m\dot{\bm\zeta}_t-e\bm A(\mathsf{R}_t\hat{\bm r}+\bm\zeta_t,t) \big]^2
% \nl
%
% &\quad-m\dot{\bm{\zeta}}_t^2/2
+m\ddot{\boldsymbol{\zeta}}_t\cdot(\mathsf{R}_t\hat{\bm r}).
\end{align}
where $\hat{\bm x}_t\equiv \mathsf{R}_t\hat{\bm r}+\bm\zeta_t$. 
See detailed derivations in Appendix A.

%When in absence of electromagnetic fields, 
%\begin{align}
%H'&=\frac{\hat{\bm p}^2}{2m}
%
%
%+V_0(\hat{\bm r})
%+m\ddot{\boldsymbol{\zeta}}_t\!\cdot\!(\mathsf{R}_t\hat{\bm r})
%\nl & \quad
%- \Omega(t){\bf n}(t)\!\cdot\!\big[(\mathsf{R}_t\hat{\bm r})\!\times\!(\mathsf{R}_t\hat{\bm p})\big],
%
%\end{align}
%which gives rise to the equation of motion
%\begin{align}
%m\ddot{ \hat{\bm r}}'=
%\end{align}

\subsection{Electromagnetic field transformation}\label{EM}

The dynamics of the electromagnetic (EM) field in the non-inertial frame is governed by the covariant Maxwell's equation, 
\begin{align}\label{Maxwelleqs}
    \tensor{F}{^\mu^\nu_{;\mu}}(x) = 0 \quad\text{and}\quad F_{\mu\nu;\lambda} + F_{\lambda\mu;\nu} + F_{\nu\lambda;\mu} = 0.
\end{align}
Here we define the covariant derivative as 
\begin{subequations}\label{covd}
\begin{align}
    \tensor{V}{^\mu_{;\nu}} \equiv \tensor{V}{^\mu_{,\nu}} + \Gamma^\mu_{\lambda\nu}V^\lambda
\end{align}
and
\begin{align}
    \tensor{V}{_\mu_{;\nu}} \equiv \tensor{V}{_\mu_{,\nu}} - \Gamma^\lambda_{\mu\nu}V_\lambda
\end{align}
for any 4-vector $V^\mu$ and $V_\mu \equiv g_{\mu\nu}V^\nu$ with $g_{\mu\nu}$ being the Riemann metric. In \Eq{covd}, we defines $f_{,\mu}(x) \equiv \frac{\partial}{\partial x^\mu}f(x)$ for any function $f(x)$ and the Christoffel connection
\end{subequations}
\begin{align}
    \Gamma^\lambda_{\mu\nu} = \frac{1}{2}g^{\lambda\rho}(g_{\mu\rho,\nu} + g_{\nu\rho,\mu} - g_{\mu\nu,\rho}).
\end{align}
The Maxwell's equation [\Eq{Maxwelleqs}] enables us to introduce the 4-vector potential $A^\mu \equiv (\phi,\bm A)$, such that 
\begin{align}
    F_{\mu\nu} = A_{\nu;\mu} - A_{\mu;\nu} = A_{\nu,\mu} - A_{\mu,\nu},
\end{align}
where the second equality holds due to the symmetry $\Gamma^\lambda_{\mu\nu} = \Gamma^\lambda_{\nu\mu}$. Consequently, we have 
\begin{align}
    \tensor{A}{^{\nu}_{;\mu}^{;\mu}} - \tensor{A}{^\mu_{;\mu}_{;\nu}} + \tensor{R}{_\rho_\nu}A^\rho = 0.
\end{align}
Besides, the gauge condition is applied together to determine the dynamics of the 4-vector potential. 

We consider the  transformation as 
\begin{align}
    \wti t = t \quad\text{and}\quad \wti{\bm r} = \wti{\mathsf R}_t^{-1}({\bm r} - \wti{\bm \zeta}_t)
\end{align}
with $g_{\mu\nu}(\bm r, t) = g_{\mu\nu}(x) =  \eta_{\mu\nu}$. The 4-vector potential transforms as 
\begin{align}
    \begin{split}
        \wti A^0(\wti x) &= A^0(x),\\
        \wti{\bm A}(\wti x) &= \wti{\mathsf R}^{-1}_t\bm A(x) + \big[ \dot{\wti{\mathsf R}}_t{ }^{\!\!-1}(\bm r - \wti{\bm\zeta}_t) - \wti{\mathsf R}_t^{-1}\dot{\wti{\bm\zeta}}_t \big]A^0(x).
    \end{split}
\end{align}
Special cases include:
\begin{enumerate}[fullwidth]
\item If the field is comoving with the particle, $\ti{\mathsf{R}}_t=\mathsf{R}_t$, $\ti{\bm\zeta}_t=\bm\zeta_t$, leading to $\ti{\bm r}={\bm r}'$ [cf.\,\Eq{potential}];
\item If the field is static, $\ti{\mathsf{R}}_t=\hat{I}$, $\ti{\bm\zeta}_t={0}$, leading to $\ti{\bm r}={\bm r}$.
\end{enumerate}

In this work, we quantize the electromagnetic field under the Coulomb gauge, reading 
\begin{align}\label{phiA}
    \wti{\bm\nabla}\cdot\wti{\bm A}(\wti{\bm r},t) = 0.
\end{align}
This leads to $\wti \phi(\wti{\bm r},t) = 0$ and 
\begin{align}\label{dAl}
    \tensor{\wti{\bm A}}{_{;\mu}^{;\mu}}(\wti{\bm r},t) = 0.
\end{align}
The general solution of \Eq{dAl} allows us to quantize the vector potential as
\begin{align}\label{dAl1}
    \!\!\!\wti{\bm A}(\ti {\bm r},t)&=\!\sum_{k,s}\bar{Z}_k{\bm\epsilon}_{k}^{s}\Big( \hat a_{k}^{s}\ue^{-i\omega_kt+i{\bm k}\cdot\ti{\bm r}}+\hat a^{s \dagger}_{k}\ue^{i\omega_kt-i{\bm k}\cdot\ti{\bm r}}\Big).
\end{align}
Here, $\bar{Z}_k$ is the normalization constant, $s=1,2$ labels the
two polarizing states of photon, and ${\bm\epsilon_{\bm k}^s}$ are the
polarization vectors perpendicular to the wave vector ${\bm k}$.
Note that we can determine $\bar Z_k$ in \Eq{dAl} via the canonical quantization condition.
Substituting \Eq{dAl1} into \Eq{dAl},  we can obtain the dispersion relation $\w_{\bm k}$. 
%
% We will elaborate the procedures with the specific example in \Sec{thsec4}.
%
One may directly apply the discussions and results of this subsection to other types of environments such as phonons.

\section{Non-inertial effects in open quantum systems}\label{thsec3}

\subsection{Total Hamiltonian with non-inertial motion}

Combining the results in \Sec{thsec2}, we arrive at the total system--plus--bath composite Hamiltonian, reading 
\begin{align}\label{Hprime}
    H' = H_{\tS} + H_{\tS\E}
\end{align}
with the system Hamiltonian and the system--environment interaction being 
\begin{align}\label{Hs}
H_{\tS}=\frac{\hat{\bm p}^2}{2m}
+V_0(\hat{\bm r})-\Omega(t){\bf n}(t)\cdot \mathsf R_t\hat {\bm L}
+m\ddot{\boldsymbol{\zeta}}_t\cdot(\mathsf{R}_t\hat{\bm r})
\end{align}
and 
\be \label{Hse}
H_{\tS\E}=-e\ti{\mathsf{R}}_t^{-1}\bigg(\frac{\mathsf{R}_t \hat{\bm p} }{m}+\dot{\boldsymbol{\zeta}}_t\bigg)\cdot\wti{\bm A}\big(\ti{\mathsf{R}}_t^{-1}(\mathsf{R}_t \hat {\bm r}+\bm\zeta_t-\ti{\bm \zeta}_t)\big),
\ee
respectively. Here, we express $H'$ in the environment Hamiltonian $H_{\E}$-interaction picture and we ignore the terms with order of $e^2$. We further adopt the long wavelength approximation for the EM field, i.e., ${\bm k}\cdot \ti{\bm r}\ll 1$. This leads to 
 \begin{align}\label{A0}
    \wti{\bm A}(\ti {\bm r},t)&\simeq \sum_{k,s}\bar{Z}_k{\bm\epsilon}_{k}^{s}\Big( \hat a_{k}^{s}\ue^{-i\omega_kt}+\hat a^{s \dagger}_{k}\ue^{i\omega_kt}\Big),
\end{align}
which is independent of the coordinate. 

% Substitute \Eq{potential} into \Eq{Hprime}, 
% noting that
% \be 
% \begin{split}
% {\bm A}(\hat{\bm x}_t,t)&=\ti{\mathsf{R}}_t \bm A_0\big(\ti{\mathsf{R}}_t^{-1}(\hat{\bm x}_t-\ti{\bm \zeta}_t),t\big)
% \nl &=\ti{\mathsf{R}}_t \bm A_0\big(\ti{\mathsf{R}}_t^{-1}(\mathsf{R}_t\hat{\bm r}+\bm\zeta_t-\ti{\bm \zeta}_t)\big),
% \end{split}
% \ee
% and we obtain the total composite Hamiltonian reads in the environment Hamiltonian $H_{\E}$-interaction picture, 
% \be \label{Hprime}
% H'=H_{\tS}+H_{\tS\E}.
% \ee
% Here, the system Hamiltonian reads
% \begin{align}\label{Hs}
% H_{\tS}=\frac{\hat{\bm p}^2}{2m}
% +V_0(\hat{\bm r})-\Omega(t){\bf n}(t)\cdot \mathsf R_t\hat {\bm L}
% +m\ddot{\boldsymbol{\zeta}}_t\cdot(\mathsf{R}_t\hat{\bm r}),
% \end{align}
% the system-environment interaction reads
% \be \label{Hse}
% H_{\tS\E}=-e\ti{\mathsf{R}}_t^{-1}\bigg(\frac{\mathsf{R}_t \hat{\bm p} }{m}+\dot{\boldsymbol{\zeta}}_t\bigg)\cdot  \bm A_0\big(\ti{\mathsf{R}}_t^{-1}(\mathsf{R}_t \hat {\bm r}+\bm\zeta_t-\ti{\bm \zeta}_t)\big),
% \ee
% up to the first order of $e$,
% and the environment-field interaction part reads 

\subsection{Spectral density}
This subsection gives a comprehensive account on the settings of open quantum system and calculates the environment spectral density. By using the long wavelength approximation, \Eq{Hse} can be recast into
\begin{align}\label{Hse_re}
    H_{\tS\E}&=-\hat{\bm{Q}}(t)\cdot  \wti{\bm A}(t),
\end{align}
where we denote $\hat{\bm Q}(t)\equiv e\ti{\mathsf{R}}_t^{-1}\Big(\frac{\mathsf{R}_t \hat{\bm p} }{m}+\dot{\boldsymbol{\zeta}}\Big)$. The environmental influence on the system is fully characterized by the autocorrelation functions of the vector potential:
\be\label{Fcorr_boson}
\begin{split}
C_{ ij}(t)\equiv \la{ \wti A}_{i}(t){\wti A}_{j}(0)\ra_{\E},
\end{split}
\ee
where $\wti A_{i}$ represents the $i$-component of $\wti{\bm A}$ in a particular spatial coordinate system. For example, $i=r$, $\theta$, $z$ in the cylindrical coordinate system. Here, $\la(\,\cdot\,)\ra_{\E}\equiv{\rm tr}_{\E}[(\,\cdot\,)
\rho^{0}_{\E}] = {\rm tr}_{\E}[(\cdot)e^{-\beta h_{\E}}/Z_{\E}]$ with $Z_{\E} \equiv {\rm tr}_{\E}(e^{-\beta h_{\E}})$. 

Or equivalently, it can be characterized by the environment spectral density:
\begin{align}\label{Jw0}
 \!\!\!\!\!J_{ij}(\w)&\equiv \frac{1}{2}\int^{\infty}_{-\infty}\ud t\, e^{i\w t}
   \la [\wti{A}_{i}(t),\wti{A}_{j}(0)]\ra_{\E}
 \nl &
 =  \pi\sum_{ks}\bar{Z}_k^2{\epsilon}_{ki}^{s}{\epsilon}_{k  j}^{s}\Big[\delta(\omega-\omega_k)-\delta(\omega+\omega_k)\Big].
\end{align} 
Here, $\la(\,\cdot\,)\ra_{\E}\equiv{\rm tr}_{\E}[(\,\cdot\,)
\rho^{0}_{\E}] = {\rm tr}_{\E}[(\cdot)e^{-\beta h_{\E}}/Z_{\E}]$ with $Z_{\E} \equiv {\rm tr}_{\E}(e^{-\beta h_{\E}})$. 
They satisfy the positivity relations, $J_{ii}(\w)/\w\geq 0$ and
$|J_{ij}(\w)|^2\leq J_{ii}(\w)J_{jj}(\w)$,
and the symmetry relations,
\be\label{Jab_sym}
 J^{\ast}_{ij}(\w)=-J_{ij}(-\w)=J_{ji}(\w).
\ee
Together with the detailed balance relation, one can readily
obtain  \cite{Wei12,Kle09,Yan05187}
\be\label{FDT}
 \la {\wti A}_{i}(t){\wti A}_{j}(0)\ra_{\E}
=\frac{1}{\pi}\int^{\infty}_{-\infty}\!\!
  \ud\w \frac{e^{-i\w t} J_{ij}(\w)}{1-e^{-\beta\w}}.
\ee
This is the bosonic fluctuation--dissipation theorem, which
relates the bath correlation function to
the spectral density function.

According to expression of $H_{\tS\E}^{(\rm II)}(t)$, the bath  can be seen as driven by an external classical field, $\ti{\mathsf{R}}_t^{-1}\dot{\boldsymbol{\zeta}}$. 
%
% The bath--field interaction Hamiltonian reads
% %
% \be 
% H^{(\rm I)}(t)=\varepsilon(t)\sum_j d_j x_j\ \ \,\text{(linearly)},
% \ee

\subsection{Dissipaton theory: An exact framework}
The HEOM starts with an exponential expansion of \Eq{Fcorr_boson},
\be\label{FF_exp}
 C_{ij}(t)
 = \sum_{\kappa=1}^{K} \eta_{ij\kappa} e^{-\gamma_{\kappa} t},
\ee
where we set $\gamma_{ij\kappa}=\gamma_\kappa$ for simplicity. This can generally be achieved via certain sum--over--poles expansion on the Fourier integrand of \Eq{FDT}, followed by the Cauchy's contour integration in the low--half plane. 

From the definition of \Eq{Fcorr_boson}, we obtain the Time-reversal relation,
\begin{align}\label{FF_exp_rev}
    C_{ji}(-t)=[C_{ij}(t)]^*=\sum_{\kappa=1}^{K} \eta_{ij\kappa}^* e^{-\gamma_{\kappa}^* t}=\sum_{\kappa=1}^{K} \eta_{ij\bar{\kappa}}^* e^{-\gamma_{\bar{\kappa}} t}.
\end{align}

 {\color{black}The DEOM theory explicitly identifies all involved dynamical variables,}
$\{\rho^{(n)}_{\textbf{n}}(t)\}$,
as follows. First of all, according to the Gaussian--Wick's thermodynamics
theorem \cite{Yan05187,Wei12}, {\color{black}the influences of the linearly coupled bath environment}
are completely characterized by the bare bath correlation
functions, \Eqs{FF_exp} and (\ref{FF_exp_rev}).
Consider now the dissipaton decomposition
on the hybridization bath operator \cite{Yan14054105,Yan16110306},
\be\label{F_in_f}
 \wti A_{i} = \sum_{\kappa=1}^K \hat f_{i\kappa}.
\ee
The involving dissipatons that recover \Eqs{FF_exp}
and (\ref{FF_exp_rev})
are statistically independent quasi--particles,
with \cite{Yan14054105,Yan16110306}
\be\label{ff_corr}
\begin{split}
 \la \hat f_{i\kappa}(t)\hat f_{j\kappa'}(0)\ra_{\B}
 &= \delta_{\kappa\kappa'}\eta_{ij\kappa}e^{-\gamma_\kappa t},
\\
 \la \hat f_{j\kappa'}(0)\hat f_{i\kappa}(t)\ra_{\B}
 &= \delta_{\kappa\kappa'}\eta^{\ast}_{ij{\bar \kappa}}e^{-\gamma_\kappa t}.
\end{split}
\ee
%%%
The above expressions, where $t>0$, highlight the
basic feature of dissipatons, whose forward and
backward correlations functions in the bare bath
ensemble share a common exponent.
This feature leads to
the \emph{generalized diffusion equation} \cite{Yan14054105,Yan16110306},
\be\label{diff}
 {\rm tr}_{\B}\Big[\Big(\frac{\partial}{\partial t} \hat f_{i\kappa}\Big)_{\B}\rho_{\T}(t)\Big]
 =-\gamma_{\kappa}\,
    {\rm tr}_{\B}\big[\hat f_{i\kappa}\rho_{\T}(t)\big] .
\ee
While $\gamma_{\kappa}$ can be complex,
the total composite $\rho_{\T}(t)$ is non-Gaussian in general.
%%%
% {\color{black} In the DEOM construction below},
% \Eq{diff} will be used together with the Heisenberg equation
% of motion in bare bath,
% \be\label{hB_Heisenberg}
%   \Big(\frac{\partial}{\partial t} \hat f_{ak}\Big)_{\B}=-i[\hat f_{ak},h_{\B}].
% \ee

 The dynamical variables in DEOM are called the dissipaton density operators (DDOs),
defined as \cite{Yan14054105,Yan16110306}:
\be\label{DDO}
 \rho^{(n)}_{\textbf{n}}(t)\equiv {\rm tr}_{\B}\Big[
  \Big(\prod_{i\kappa} \hat f^{n_{i\kappa}}_{i\kappa}\Big)^\circ
  \rho_{\T}(t)\Big].
\ee
{\color{black}
Here, $n=\sum_{i\kappa} n_{i\kappa}$ and $\textbf{n}=\{n_{i\kappa}\}$
that is an ordered set of the occupation numbers, $n_{i\kappa}=0,1,\cdots$,
on individual dissipatons.
The circled parentheses, $(\cdots)^{\circ}$, is \emph{irreducible} notation,
so that all the $c$-numbers in the normal ordering of dissipatons product vanish. 
For bosonic dissipatons it follows that
$(\hat f_{i\kappa}\hat f_{j\kappa'})^{\circ}=(\hat f_{j\kappa'}\hat f_{i\kappa})^{\circ}$.
In other words, the irreducible product of dissipatons
inside $(\cdots)^{\circ}$ in \Eq{DDO} resembles the
second--quantization representation of a bosonic permanent.}
%%%
{\color{black} The DDOs for fermionic coupled environment are similar,
but resemble a Slater determinant,}
having the occupation number of $0$ or $1$ only,
due to the antisymmetric permutation relation \cite{Yan14054105,Yan16110306}.
%%%%
Therefore, $\rho^{(n)}_{\textbf{n}}(t)$ of \Eq{DDO}
specifies an $n$--dissipaton configuration,
with $\rho_{\bf 0}^{(0)}(t)=\rho_{\tS}(t)$ being just the
reduced system density operator.
%%%%%
Denote also $\rho^{(n\pm 1)}_{{\bf n}^{\pm}_{i\kappa}}$ as the associated
$(n\pm 1)$-dissipatons configuration, with
${\bf n}^{\pm}_{ik}$ differing from ${\bf n}$ only
at the specified $\hat f_{ik}$-disspaton occupation number,
$n_{i\kappa}$, by $\pm 1$.

 The most important ingredient in the dissipaton algebra
is the \emph{generalized Wick's theorem} \cite{Yan14054105,Yan16110306}:
\begin{align}\label{Wick1}
 &\quad {\rm tr}_{\B}\Big[\Big(\prod_{i\kappa} \hat f^{n_{i\kappa}}_{i\kappa}\Big)^\circ
   \hat f_{j\kappa'} \rho_{\T}(t)\Big]
\nl&=
 \rho^{(n+1)}_{{\bf n}^{+}_{j\kappa'}}(t)+\sum_{i\kappa} n_{i\kappa}\la\hat f_{i\kappa}\hat f_{j\kappa'}\ra^{>}_{\B}
   \rho^{(n-1)}_{{\bf n}^{-}_{i\kappa}}(t),
\end{align}
and
\begin{align}\label{Wick2}
 &\quad {\rm tr}_{\B}\Big[\Big(\prod_{i\kappa} \hat f^{n_{i\kappa}}_{i\kappa}\Big)^\circ
   \rho_{\T}(t)\hat f_{j\kappa'} \Big]
\nl&=
 \rho^{(n+1)}_{{\bf n}^{+}_{j\kappa'}}(t)+\sum_{i\kappa} n_{i\kappa}\la\hat f_{jk'}\hat f_{i\kappa}\ra^{<}_{\B}
   \rho^{(n-1)}_{{\bf n}^{-}_{i\kappa}}(t).
\end{align}
{\color{black} The involved forward $\la\hat f_{i\kappa}\hat f_{j\kappa'}\ra^{>}_{\B}$
and backward $\la\hat f_{j\kappa'}\hat f_{i\kappa}\ra^{<}_{\B}$ coefficients}
are related to the correlation
functions in \Eq{ff_corr} as
\be\label{ff0}
\begin{split}
 \la\hat f_{i\kappa}\hat f_{j\kappa'}\ra^{>}_{\B}
 &\equiv \la\hat f_{i\kappa}(0+)\hat f_{j\kappa'}(0)\ra_{\B} = \eta_{ij\kappa}\delta_{\kappa\kappa'},
\\
 \la\hat f_{j\kappa'}\hat f_{i\kappa}\ra^{<}_{\B}
 &\equiv \la\hat f_{j\kappa'}(0)\hat f_{i\kappa}(0+)\ra_{\B} = \eta^{\ast}_{ij{\bar \kappa}}\delta_{\kappa\kappa'}.
\end{split}
\ee

 The DEOM can now be readily constructed
by applying $\dot{\rho}_{\T}(t)=-i[H'(t),\rho_{\T}(t)]$,
{\color{black} to the total composite density operator in \Eq{DDO};}
i.e.,
\be\label{DDO_dot}
 \dot\rho^{(n)}_{\textbf{n}}(t)= -i\, {\rm tr}_{\B}\Big\{
  \Big(\prod_{i\kappa} \hat f^{n_{i\kappa}}_{i\kappa}\Big)^\circ
  [H'(t),\rho_{\T}(t)]\Big\}.
\ee
To proceed, we express the total composite Hamiltonian,
\Eq{Hprime} with \Eqs{Hs}, (\ref{Hse}), (\ref{Hse_re}) and (\ref{F_in_f}), as
\be\label{HT}
  H'(t)=H_{\tS}(t)+\sum_{i\kappa}  \hat Q_{i}(t)\hat f_{i\kappa}.
\ee
Equation (\ref{DDO_dot}) is then evaluated by using \Eq{diff}
% with \Eq{hB_Heisenberg} for the action of $h_{\B}$
, and \Eqs{Wick1}--(\ref{ff0})
for the action of the last term in \Eq{HT}.
We obtain \cite{Yan14054105,Yan16110306}
\begin{align}\label{DEOM}
 \dot\rho^{(n)}_{\bf n}= & -i [H_{\tS}(t),\rho^{(n)}_{\bf n}] -\sum_{i\kappa} n_{i\kappa}\gamma_\kappa\rho^{(n)}_{\bf n} -i\sum_{i\kappa} [ Q_{i}(t),\rho^{(n+1)}_{{\bf n}_{i\kappa}^+}] \nl&
-i\sum_{ij\kappa}n_{i\kappa}\Big[\eta_{ij\kappa}\hat Q_{j}(t)\rho^{(n-1)}_{{\bf n}_{i\kappa}^-} -\eta^{\ast}_{ij\bar \kappa}\rho^{(n-1)}_{{\bf n}_{i\kappa}^-}\hat Q_j(t) \Big].
\end{align}
This is a temperature--independent real parameter.
{\color{black} The contributing coefficients, $\eta_{ij\kappa}$ and $\eta^{\ast}_{ij{\bar \kappa}}$, arise
solely from the poles of the bath spectral density;
}
see the comments after \Eq{FF_exp}.
% Those from the Bose function
% are of real $\gamma_{k}$ and $\eta^{\ast}_{ab{\bar k}}=\eta^{\ast}_{abk}
% =\eta_{abk}$, having no contributions.

\section{Summary}\label{thsecsum}

This work presents a universal formalism of open system quantum mechanics with non-inertial motions. We establish the theory based on a charged system interacting with the electromagnetic field environment. The formulation allows the non-inertial motions of the system and environment to be different. Under the non-inertial unitary transformation, the electromagnetic field still satisfies the Gauss--Wick's statistics, leading to the validity of the dissipaton decomposition of the vector potential $\wti{\bm A}$. The corresponding dissipaton equation of motion presents an exact numerical approach to the reduced system dynamics and correlations \cite{Yan14054105,Yan16110306}. Further developments may include extending the formalism to the relativistic Poincar\'e transformation and applying to exploring the non-inertial effects on open quantum systems, including the cavity QED and chemical reactions.

\begin{acknowledgments}
Support from
the Ministry of Science and Technology of China (Grant No.\  2021YFA1200103), the National Natural Science Foundation of China (Grant Nos.\ 22103073,  22173088, and 22373091) and the Innovation Program for Quantum Science and Technology
(Grant No.\ 2021ZD0303301) is gratefully acknowledged.
 \end{acknowledgments}

\appendix
\section{Derivation of \Eq{Hprime}}
Firstly, we rewrite the unitary transformation in \Eq{uni} as
\be \label{unitary}
U(t)=\exp\bigg(\!-im\!\int_{0}^t\!\ud\tau\,\frac{\dot{\boldsymbol{\zeta}}_{\tau}^2}{2}\bigg)\cdot
U_1(t)\cdot U_{2}(t),
\ee
where
\bsube \label{U1U2}
\be 
U_1(t)=\exp_{-}\bigg[i\int_{0}^{t}\!\ud\tau\, \Omega(\tau){\bf n}(\tau)\cdot\hat{\bm L}\bigg],
\ee
and
\be 
U_2(t)=
   \exp\Big(-im\dot{\boldsymbol{\zeta}}_t\cdot\hat{\bm r}\Big) \cdot\exp\Big(i\boldsymbol{\zeta}_t\cdot\hat{\bm p}\Big).
\ee 
\esube

Consider the $UHU^{\dg}$ term. Evidently, 
\begin{align}
U(t)HU^{\dg}(t)=U_1(t)U_2(t)HU_2^{\dg}(t)U_1^{\dg}(t), 
\end{align}
since $\exp\big(\!-im\!\int_{0}^t\!\ud\tau\,\dot{\boldsymbol{\zeta}}_{\tau}^2/2\big)$ is a pure phase factor. Then by using
\bsube\label{U1}
\begin{align}
U_2(t)\hat{\bm r} U_2^{\dg}(t)&=\hat{\bm r}+\bm\zeta_t,
\\
U_2(t)\hat{\bm p} U_2^{\dg}(t)&=\hat{\bm p}+m\dot{\bm\zeta}_t,
\end{align}
\esube
we obtain
\begin{align}\label{first}
U(t)HU^{\dg}(t)=U_1(t)H(\hat{\bm r}+\bm\zeta_t,\hat{\bm p}+m\dot{\bm\zeta}_t)U_1^{\dg}(t). 
\end{align}
Further noting 
\bsube\label{U2}
\begin{align}
    U_1(t)(\hat{\bm r}+\bm\zeta_t )U_1^{\dagger}(t)=\mathsf{R}_t\hat{\bm r}+\bm\zeta_t,
    \\
      U_1(t)(\hat{\bm p}+m\dot{\bm\zeta}_t)U_1^{\dagger}(t)=\mathsf{R}_t\hat{\bm p}+m\dot{\bm\zeta}_t,
\end{align}
\esube
we have
\begin{align}
U(t)HU^{\dg}(t)
%&=H(\mathsf{R}_t\hat{\bm r}+\bm\zeta_t,\,\mathsf{R}_t\hat{\bm p}+m\dot{\bm\zeta}_t)
%
%\nl
%
&=\frac{1}{2m}\big[\mathsf{R}_t\hat{\bm p}+m\dot{\bm\zeta}_t-e\bm A(\mathsf{R}_t\hat{\bm r}+\bm\zeta_t,t) \big]^2
\nl
&\quad+V_0(\hat{\bm r})+e\phi(\mathsf{R}_t\hat{\bm r}+\bm\zeta_t,t).
\end{align}

For the $i\dot{U}U^{\dagger}$ term, according to \Eq{unitary}, we have
\begin{align}\label{second}
   i\dot{U}U^{\dagger}
   &=-m\dot{\bm{\zeta}}_t^2/2
+m\ddot{\boldsymbol{\zeta}}_t\cdot(\mathsf{R}_t\hat{\bm r})
-\dot{\bm{\zeta}}_t\cdot(\mathsf{R}_t\hat{\bm p})
\nl
&\quad- \Omega(t){\bf n}(t)\cdot\big[(\mathsf{R}_t\hat{\bm r})\times(\mathsf{R}_t\hat{\bm p})\big], 
\end{align}
where we have used \Eqs{U1} and (\ref{U2}) with noting that
$U_1\hat{\bm L}U^{\dagger}_1=(\mathsf{R}_t\hat{\bm r})\times(\mathsf{R}_t\hat{\bm p}) = \mathsf R_t\hat {\bm L}$. The combination of \Eqs{first} and (\ref{second}) gives rise to the expression of \Eq{Hprime}.

%   \bibliographystyle{./aiptit}
% \bibliography{./bibrefs}

\begin{thebibliography}{10}

    \bibitem{Wei21}
    U.~Weiss,
    \newblock {\em Quantum Dissipative Systems},
    \newblock World Scientific, Singapore, 2021,
    \newblock 5th edition.
    
    \bibitem{Kle09}
    H.~Kleinert,
    \newblock {\em Path Integrals in Quantum Mechanics, Statistics, Polymer
      Physics, and Financial Markets},
    \newblock World Scientific, Singapore, 5th edition, 2009.
    
    \bibitem{Bre02}
    H.~P. Breuer and F.~Petruccione,
    \newblock {\em The Theory of Open Quantum Systems},
    \newblock Oxford University Press, New York, 2002.
    
    \bibitem{Bre16021002}
    H.-P. Breuer, E.-M. Laine, J.~Piilo, and B.~Vacchini, \newblock ``Colloquium:
      Non-Markovian dynamics in open quantum systems,'' Rev. Mod. Phys. {\bf 88},
      021002 (2016).
    
    \bibitem{Dev17015001}
    I.~de~Vega and D.~Alonso, \newblock ``Dynamics of non-Markovian open quantum
      systems,'' Rev. Mod. Phys. {\bf 89}, 015001 (2017).
    
    \bibitem{Scu97}
    M.~O. Scully and M.~S. Zubairy,
    \newblock {\em Quantum Optics},
    \newblock Cambridge University Press, Cambridge, 1997.
    
    \bibitem{Lou73}
    W.~H. Louisell,
    \newblock {\em Quantum Statistical Properties of Radiation},
    \newblock Wiley, New York, 1973.
    
    \bibitem{Haa7398}
    F.~Haake, \newblock ``Statistical treatment of open systems by generalized
      master equations,'' in {\em Quantum Statistics in Optics and Solid State
      Physics: Springer Tracts in Modern Physics, Vol.~66}, edited by
      G.~{H\"{o}hler}, pages 98--168, Springer, Berlin, 1973.
    
    \bibitem{Hak70}
    H.~Haken,
    \newblock {\em Laser Theory},
    \newblock Springer, Berlin, 1970.
    
    \bibitem{Sar74}
    M.~{Sargent~III}, M.~O. Scully, and J.~W.~E.~Lamb,
    \newblock {\em Laser Physics},
    \newblock Addison-Wesley, Reading, MA, 1974.
    
    \bibitem{Rei82}
    P.~Reineker,
    \newblock {\em Exciton Dynamics in Molecular Crystals and Aggregates:
      Stochastic Liouville Equation Approach: Coupled Coherent and Incoherent
      Motion, Optical Line Shapes, Magnetic Resonance Phenomena},
    \newblock Springer, Berlin, 1982.
    
    \bibitem{Sli90}
    C.~P. Slichter,
    \newblock {\em Principles of Magnetic Resonance},
    \newblock Springer Verlag, New York, 1990.
    
    \bibitem{Van051037}
    L.~M.~K. Vandersypen and I.~L. Chuang, \newblock ``NMR techniques for quantum
      control and computation,'' Rev. Mod. Phys. {\bf 76}, 1037 (2005).
    
    \bibitem{Bor85}
    M.~Born and K.~Huang,
    \newblock {\em Dynamical Theory of Crystal Lattices},
    \newblock Oxford University Press, New York, 1985.
    
    \bibitem{Hol59325}
    T.~Holstein, \newblock ``Studies of polaron motion {Part I.\ The}
      molecular-crystal model,'' Ann. Phys. {\bf 8}, 325 (1959).
    
    \bibitem{Hol59343}
    T.~Holstein, \newblock ``Studies of polaron motion {Part II.\ The} ``small''
      polaron,'' Ann. Phys. {\bf 8}, 343 (1959).
    
    \bibitem{Kli97}
    C.~F. Klingshirn,
    \newblock {\em Semiconductor Optics},
    \newblock Springer-Verlag, Heidelberg, 1997.
    
    \bibitem{Ram98}
    J.~Rammer,
    \newblock {\em Quantum Transport Theory},
    \newblock Perseus Books, Reading, Mass., 1998.
    
    \bibitem{Aka15056002}
    Y.~Akamatsu, \newblock ``Heavy quark master equations in the Lindblad form at
      high temperatures,'' Phys. Rev. D {\bf 91}, 056002 (2015).
    
    \bibitem{Bla181}
    J.-P. Blaizot and M.~A. Escobedo, \newblock ``Quantum and classical dynamics of
      heavy quarks in a quark-gluon plasma,'' J. High Energy Phys. {\bf 2018}, 1
      (2018).
    
    \bibitem{Miu20034011}
    T.~Miura, Y.~Akamatsu, M.~Asakawa, and A.~Rothkopf, \newblock ``Quantum
      Brownian motion of a heavy quark pair in the quark-gluon plasma,'' Phys. Rev.
      D {\bf 101}, 034011 (2020).
    
    \bibitem{Yao212130010}
    X.~Yao, \newblock ``Open quantum systems for quarkonia,'' Int. J. Mod. Phys. A
      {\bf 36}, 2130010 (2021).
    
    \bibitem{Muk95}
    S.~Mukamel,
    \newblock {\em The Principles of Nonlinear Optical Spectroscopy},
    \newblock Oxford University Press, New York, 1995.
    
    \bibitem{She84}
    Y.~R. Shen,
    \newblock {\em The Principles of Nonlinear Optics},
    \newblock Wiley, New York, 1984.
    
    \bibitem{Muk81509}
    S.~Mukamel, \newblock ``Reduced equations of motion for collisionless molecular
      multiphoton processes,'' Adv. Chem. Phys. {\bf 47}, 509 (1981).
    
    \bibitem{Yan885160}
    Y.~J. Yan and S.~Mukamel, \newblock ``Electronic dephasing, vibrational
      relaxation, and solvent friction in molecular nonlinear optical lineshapes,''
      J. Chem. Phys. {\bf 89}, 5160 (1988).
    
    \bibitem{Yan91179}
    Y.~J. Yan and S.~Mukamel, \newblock ``Photon echoes of polyatomic molecules in
      condensed phases,'' J. Chem. Phys. {\bf 94}, 179 (1991).
    
    \bibitem{Che964565}
    V.~Chernyak and S.~Mukamel, \newblock ``Collective coordinates for nuclear
      spectral densities in energy transfer and femtosecond spectroscopy of
      molecular aggregates,'' J. Chem. Phys. {\bf 105}, 4565 (1996).
    
    \bibitem{Tan939496}
    Y.~Tanimura and S.~Mukamel, \newblock ``Two-dimensional femtosecond vibrational
      spectroscopy of liquids,'' J. Chem. Phys. {\bf 99}, 9496 (1993).
    
    \bibitem{Tan943049}
    Y.~Tanimura and S.~Mukamel, \newblock ``Multistate quantum Fokker-Planck
      approach to nonadiabatic wave packet dynamics in pump-probe spectroscopy,''
      J. Chem. Phys. {\bf 101}, 3049 (1994).
    
    \bibitem{Nit06}
    A.~Nitzan,
    \newblock {\em Chemical Dynamics in Condensed Phases: Relaxation, Transfer and
      Reactions in Condensed Molecular Systems},
    \newblock Oxford University Press, New York, 2006.
    
    \bibitem{Lee071462}
    H.~Lee, Y.-C. Cheng, and G.~R. Fleming, \newblock ``Coherence dynamics in
      photosynthesis: Protein protection of excitonic coherence,'' Science {\bf
      316}, 1462 (2007).
    
    \bibitem{Eng07782}
    G.~S. Engel, T.~R. Calhoun, E.~L. Read, T.~K. Ahn, T.~Man\v{c}al, Y.~C. Cheng,
      R.~E. Blankenship, and G.~R. Fleming, \newblock ``Evidence for wavelike
      energy transfer through quantum coherence in photosynthetic systems,'' Nature
      {\bf 446}, 782 (2007).
    
    \bibitem{Dor132746}
    K.~E. Dorfman, D.~V. Voronine, S.~Mukamel, and M.~O. Scully, \newblock
      ``Photosynthetic reaction center as a quantum heat engine,'' Proc. Natl.
      Acad. Sci. {\bf 110}, 2746 (2013).
    
    \bibitem{Cre13253601}
    C.~Creatore, M.~A. Parker, S.~Emmott, and A.~W. Chin, \newblock ``Efficient
      biologically inspired photocell enhanced by delocalized quantum states,''
      Phys. Rev. Lett. {\bf 111}, 253601 (2013).
    
    \bibitem{Kun22015101}
    S.~Kundu, R.~Dani, and N.~Makri, \newblock ``B800-to-B850 relaxation of
      excitation energy in bacterial light harvesting: All-state, all-mode path
      integral simulations,'' J. Chem. Phys. {\bf 157}, 015101 (2022).
    
    \bibitem{Max191}
    M.~Schlosshauer, \newblock ``Quantum decoherence,'' Phys. Rep. {\bf 831}, 1
      (2019).
    
    \bibitem{Gho14527}
    M.~Ghorashi, S.A.A.and~Aminjavaheri and M.~Bagheri~Harouni, \newblock ``Quantum
      decoherence of Dirac fields in non-inertial frames beyond the single-mode
      approximation,'' Quantum Inf Process {\bf 13}, 527–545 (2014).
    
    \bibitem{Yan15321}
    Y.~Yu and L.~Ye, \newblock ``Protecting entanglement from amplitude damping in
      non-inertial frames by weak measurement and reversal,'' Quantum Information
      Processing {\bf 14}, 321 (2015).
    
    \bibitem{Ahm144996}
    M.~Ahmadi, D.~E. Bruschi, and G.~Adesso, \newblock ``Relativistic Quantum
      Metrology: Exploiting relativity to improve quantum measurement
      technologies,'' Scientific Reports {\bf 4}, 4996 (2014).
    
    \bibitem{Li15Arxiv1507_08790}
    S.-W. Li, Z.~H. Wang, L.~Zhou, and C.~P. Sun, \newblock ``Quantum optics in a
      non-inertial reference frame: the Rabi splitting in a rotating ring cavity,''
      arXiv:1507.08790  (2015).
    
    \bibitem{Ary23085011}
    N.~Arya and S.~K. Goyal, \newblock ``Lamb shift as a witness for quantum
      noninertial effects,'' Phys. Rev. D {\bf 108}, 085011 (2023).
    
    \bibitem{Hor09865}
    R.~Horodecki, P.~Horodecki, M.~Horodecki, and K.~Horodecki, \newblock ``Quantum
      entanglement,'' Rev. Mod. Phys. {\bf 81}, 865 (2009).
    
    \bibitem{Sun1790}
    W.-Y. Sun, D.~Wang, J.~Yang, and L.~Ye, \newblock ``Enhancement of multipartite
      entanglement in an open system under non-inertial frames,'' Quantum
      Information Processing {\bf 16}, 90 (2017).
    
    \bibitem{Ren1770}
    J.~G. Ren et~al., \newblock ``Ground-to-satellite quantum teleportation,''
      Nature {\bf 549}, 70 (2017).
    
    \bibitem{Pir201012}
    S.~Pirandola, U.~L. Andersen, L.~Banchi, M.~Berta, D.~Bunandar, R.~Colbeck,
      D.~Englund, T.~Gehring, C.~Lupo, C.~Ottaviani, J.~L. Pereira, M.~Razavi,
      J.~S. Shaari, M.~Tomamichel, V.~C. Usenko, G.~Vallone, P.~Villoresi, and
      P.~Wallden, \newblock ``Advances in quantum cryptography,'' Adv. Opt. Photon.
      {\bf 12}, 1012 (2020).
    
    \bibitem{Met14771}
    N.~Metwally and A.~Sagheer, \newblock ``Quantum coding in non-inertial
      frames,'' Quantum Information Processing {\bf 13}, 771–780 (2014).
    
    \bibitem{Yan14054105}
    Y.~J. Yan, \newblock ``Theory of open quantum systems with bath of electrons
      and phonons and spins: Many-dissipaton density matrixes approach,'' J. Chem.
      Phys. {\bf 140}, 054105 (2014).
    
    \bibitem{Wan22170901}
    Y.~Wang and Y.~J. Yan, \newblock ``Quantum mechanics of open systems:
      Dissipaton theories,'' J. Chem. Phys. {\bf 157}, 170901 (2022).
    
    \bibitem{Tak91463}
    S.~Takagi, \newblock ``{Quantum dynamics and non-Inertial frames of reference.
      I: Generality},'' Prog. Theor. Phys. {\bf 85}, 463 (1991).
    
    \bibitem{Wei12}
    U.~Weiss,
    \newblock {\em Quantum Dissipative Systems},
    \newblock World Scientific, Singapore, 2012,
    \newblock 4$^{\rm rd}$ ed.
    
    \bibitem{Yan05187}
    Y.~J. Yan and R.~X. Xu, \newblock ``Quantum mechanics of dissipative systems,''
      Annu. Rev. Phys. Chem. {\bf 56}, 187 (2005).
    
    \bibitem{Yan16110306}
    Y.~J. Yan, J.~S. Jin, R.~X. Xu, and X.~Zheng, \newblock ``Dissipaton equation
      of motion approach to open quantum systems,'' Frontiers Phys. {\bf 11},
      110306 (2016).
    
\end{thebibliography}

\end{document}